\pdfoutput=1
\documentclass[journal]{IEEEtran}
\ifCLASSINFOpdf
\else
  \usepackage[dvips]{graphicx}
\fi

\usepackage{graphicx,amsmath,subfigure,epstopdf,color}
\usepackage{amsmath,graphicx,tabularx}
\usepackage{color,subfigure}
\usepackage{tikz}
\usetikzlibrary{intersections}
\usepackage{graphicx,amsmath,multirow}
\usepgflibrary{arrows} 
\usepgflibrary[arrows] 
\usetikzlibrary{arrows} 
\usetikzlibrary[arrows] 

\usepackage{fancyhdr}
\renewcommand{\footskip}{20 pt}

\begin{document}
%

\title{Self-consistent Modeling of the $I_c$ of HTS Devices: How Accurate do Models Really Need to Be?}


%
%
%

\author{Francesco~Grilli, Fr\'ed\'eric Sirois~\IEEEmembership{Member,~IEEE}, Victor M. R. Zerme\~no, Michal~Vojen\v ciak
\thanks{F. Grilli and V. Zerme\~no are with the Karlsruhe Institute of Technology. F. Sirois is with the Ecole Polytechnique Montr\'eal, Montr\'eal, Canada. M. Vojen\v ciak is with the Slovak Academy of Science, Bratislava, Slovakia.}
\thanks{Funding from the following sources is gratefully acknowledged: Helmholtz-University Young Investigator Group Grant VH-NG- 617; Natural Sciences and Engineering Research Council of Canada  (NSERC).}
\thanks{Manuscript received \today}}

\maketitle

\begin{abstract}
Numerical models for computing the effective critical current of devices made of HTS tapes require the  knowledge of the $J_c(B,\theta)$ dependence, i.e. of the way the critical current density $J_c$ depends on the magnetic flux density $B$ and its orientation $\theta$ with respect to the tape.  
In this paper we present a numerical model based on the critical state with angular field dependence of $J_c$ to extract the $J_c(B,\theta)$ relation from experimental data. The model takes into account the self-field created by the tape, which gives an important contribution when the field applied in the experiments is low.

The same model can also be used to compute the effective critical current of devices composed of electromagnetically interacting tapes. In this work, we consider three examples: two differently current rated Roebel cables composed of ten strands from {\it RE}BCO coated conductors and a power cable prototype composed of 22 Bi-2223 tapes. The critical currents computed with the numerical model show good agreement with the measured ones. The simulations reveal also that several parameter sets in the $J_c(B,\theta)$ give an equally good representation of the experimental characterization of the tapes and that the measured $I_c$ values of cables are subjected to the influence of experimental conditions, such as $I_c$ degradation due to the manufacturing and assembling process and non-uniformity of the tape properties. These two aspects make the determination of a very precise $J_c(B,\theta)$ expression probably unnecessary, as long as that expression is able to reproduce the main features of the observed angular dependence.
The easiness of use of this model -- which can be straightforwardly implemented in finite-element programs able to solve static electromagnetic problems -- is very attractive both for researchers and devices manufactures who want to characterize superconducting tapes and calculate the effective critical current of superconducting devices.
\end{abstract}
\section{Introduction}
\IEEEPARstart{E}{lectromagnetic} numerical models of superconductors allow predicting the performance of superconducting devices with high accuracy. In order to provide a realistic description, the superconductor's properties used as input of the models must be accurately chosen, most notably the variation of the critical current density with the magnetic field: a typical example is given by {\it RE}BCO coated conductors, where pinning centers aimed at improving the superconductor's in-field behavior can produce complicated angular dependences of the critical current -- see for example~\cite{MacManus-Driscoll:NMAT04, Maiorov:APL05, Holesinger:SST09, Zhang:PhysC09, Selvamanickam:PhysC09, Selvamanickam:SST12}.

Numerical models typically need the knowledge of the dependence of the critical current density $J_c$ on the amplitude and orientation of the {\it local} magnetic flux density. Experimental data from tape characterization are often in the form of the tape's critical current $I_c$ as a function of the amplitude and orientation of the applied magnetic field $H_a$. This characterization -- consisting in current-voltage characteristics measured for different applied field -- includes the effects of the self-field caused by the transport current. In the case of {\it RE}BCO coated conductors at 77~K, this field is typically in the orders of several tens of mT and must be taken into account if one wants to extract a $J_c(B,\theta)$ function to be used in low-field applications, such as for example cables, fault current limiters and, to a lesser extent, generators and transformers.

In this work we are interested in proposing an easy-to-use model that can do two things: 
\begin{enumerate}
\item Extract the $J_c(B,\theta)$ dependence of superconductors from experimental $I_c(H_a,\theta)$ characteristics (inverse problem);
\item Starting from a known $J_c(B,\theta)$, calculate the effective critical current of a superconducting assembly composed of many interacting tapes (direct problem).
\end{enumerate}

\begin{figure}[t!]
\centering
\includegraphics[width= 8 cm ]{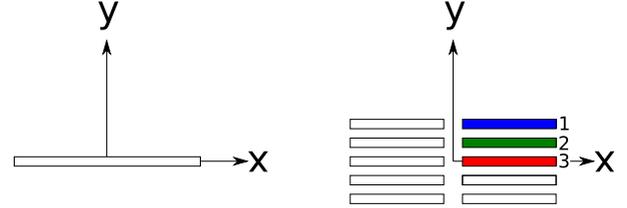}
\caption{\label{fig:drawing}{Schematic illustration of the utilized geometries: cross-section of a single tape (left) and of a Roebel cable (right). All the models are 2-D, so that the current flows perpendicular to the $x-y$ plane. The colored rectangles represent the tapes along whose width the current profiles are calculated later in the paper.}}
\end{figure}

The paper is organized as follows: in section~\ref{sec:extraction} we describe the method for extracting the parameters of the $J_c(B,\theta)$ functional dependence from experimental data. In section~\ref{sec:effIc}, starting from known $J_c(B,\theta)$ dependences,  we compute the effective $I_c$ of three different cables and we compare the results with those obtained experimentally. In the conclusion, we summarize the main findings of this work.

\section{Extraction of $J_c(B,\theta)$ dependence from experimental data}\label{sec:extraction}

\subsection{Previous work}

Several works have been dedicated to extracting the $J_c(B,\theta)$ from experimental data, which constitutes an ill-posed inverse problem. The greatest efforts in obtaining optimal curve fitting were realized by Rostila~{\it et al.} \cite{Rostila:SST07} and Sirois~{\it et al.}~\cite{Sirois:ACE02}, which both noticed that only the Nelder-Mead algorithm could lead to convergence of the optimization algorithm, despite the fact that this algorithm is known to converge to non-stationary points~\cite{McKinnon:SIAMJO98}, hence not ensuring convergence to even a local minima, least to the global minima sought. The two authors used different models for the direct problem, namely the critical state model in~\cite{Rostila:SST07}, and a static nonlinear current flow model in~\cite{Sirois:ACE02}.

Other authors such as Pardo {\it et al.} avoided entirely the optimization problem and instead proposed an intuitive way to accurately describe the complex $J_c(B,\theta)$ relationship for {\it RE}BCO coated conductors with artificial pinning centers~\cite{Pardo:SST11}: based on the position of the observed peaks in the angular dependences of $I_c$, they proposed a $J_c(B,\theta)$ consisting of three overlapping elliptical dependences, for a total of 11 parameters. The technique can provide an angular dependence of $I_c$ in excellent agreement with experimental data, given the user is skilled at finding the parameters and has a good judgement all along the manual tuning (see appendix in~\cite{Pardo:SST11}). As a consequence, the developed technique does not provide any quick means to check if other parameter sets provide an equally good fit.

Finally, Zhang {\it et al.} developed an alternative method, based on taking the experimental data of the $I_c$ reduction in parallel and perpendicular field and by using a modulating angular function $G(\theta)$ based on the measurement of $I_c$ as a function of the angle for a particular value of the applied field (100 mT in that paper)~\cite{Zhang:JAP12}. The modulating function is then used in the $J_c(B,\theta)$ dependence. The method allowed obtaining a good match for the critical current of pancake coils made of the characterized tapes.

Based on the observations above, we now address the trade-off between the need for a robust parameter identification technique vs. the accuracy provided by an optimal curve fitting. This assessment is quantified by looking at the quality of the prediction of the global critical current of assemblies of superconducting wires. In order to do this, we introduce in the next section i) a simple method to solve the direct electromagnetic problem, and ii) a brute force method that allows determining the absolute best curve fitting for a given $J_c(B,\theta)$ model. For the sake of computational efficiency and simplicity, we tried to take advantage of the most simple strategies used in previous works, namely i) by taking a static model for the direct problem, and ii) by making sure this model can be implemented in any commercial finite element software able to solve static electromagnetic problems.

\subsection{Towards a simpler approach}

Most superconducting applications can be modelled in two dimensions, by considering the 2-D transversal cross-section of the tapes ($x-y$ plane in Fig.~\ref{fig:drawing}), and assuming that there is no change in the electromagnetic properties of the tapes along their length.

\begin{figure}[t!]
\centering
\includegraphics[width=8 cm ]{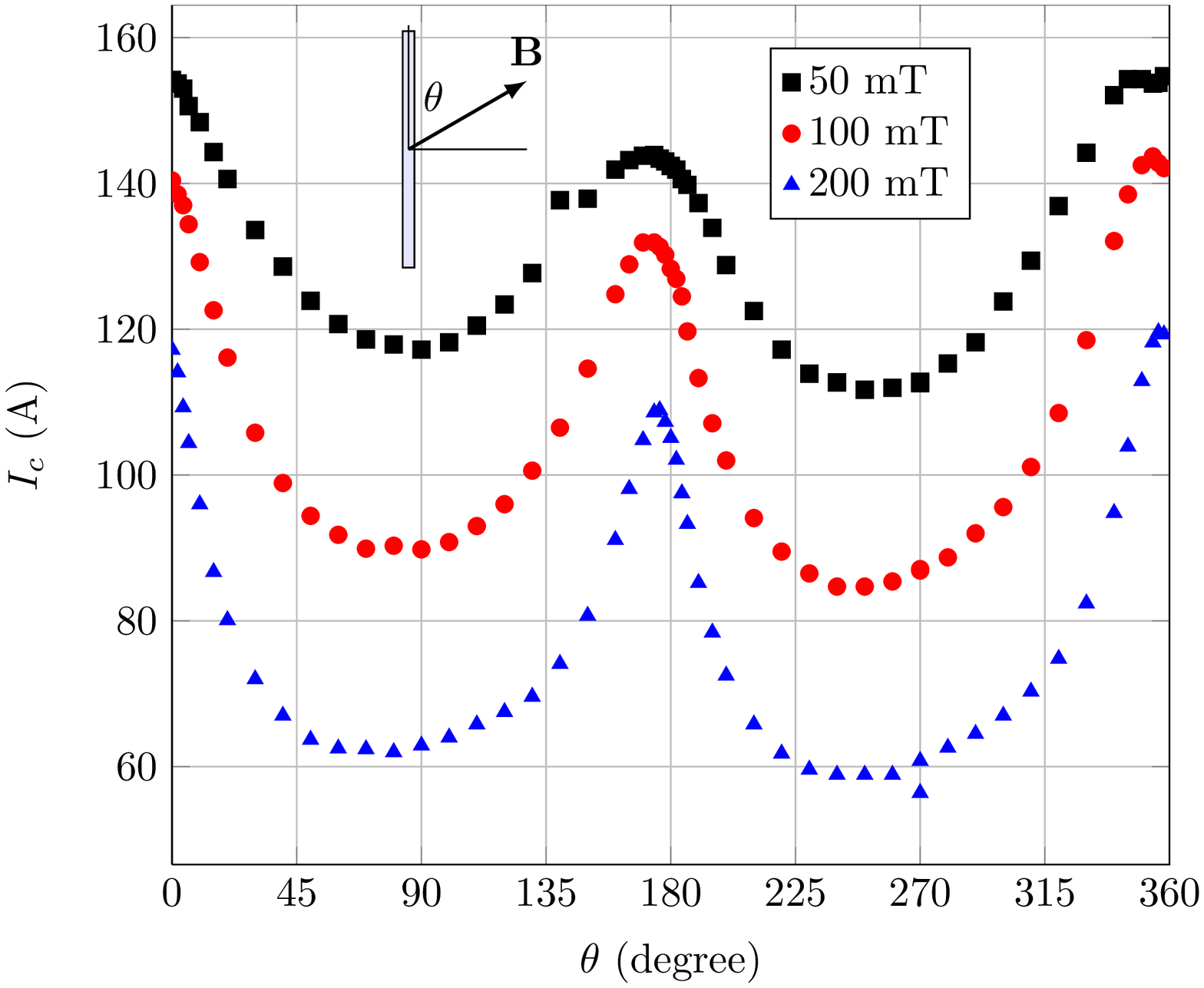}
\caption{\label{fig:IcH_exp_data_plot}{Measured angular dependence of the critical current of a single {\it RE}BCO tape.}}
\end{figure}

\begin{figure}[t!]
\centering
\includegraphics[width=8 cm ]{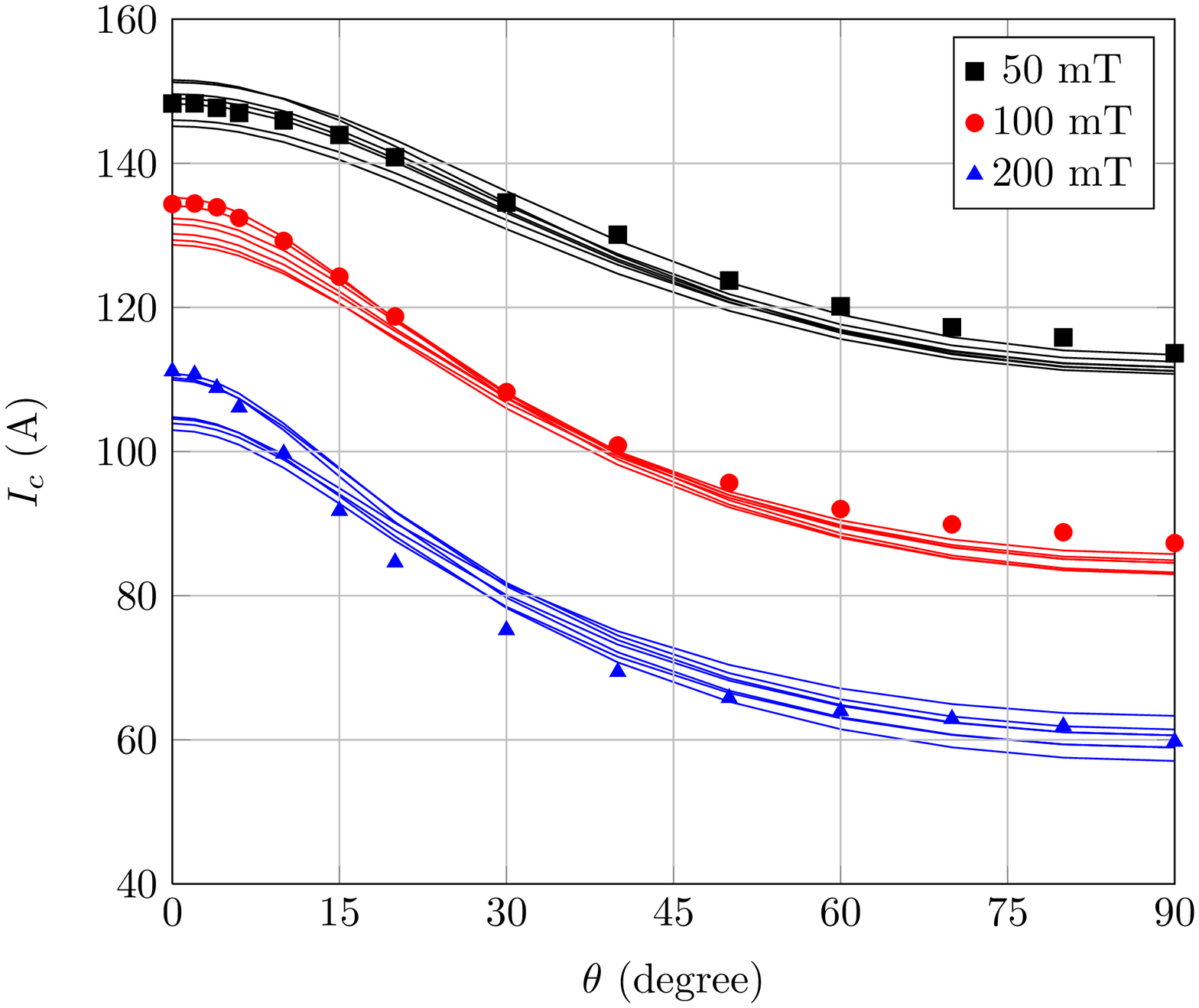}
\caption{\label{fig:best7}{Calculated angular dependence of the critical current $I_c$ for the best seven parameter sets (continuous lines) and experimental values (symbols). For the latter, the data of Fig.~\ref{fig:IcH_exp_data_plot} were averaged over the four quadrants. The self-field calculated $I_c$ values are given in the last column of Table \ref{tab:best7}.}}
\end{figure}

For this work we used 4 mm wide coated conductor tape from \emph{Superpower, Inc.}, with a self-field critical current  of 158~A at 77 K. The tape was characterized in applied magnetic field of varying amplitude and orientation, and the measured angular dependence of $I_c$ is shown in Fig.~\ref{fig:IcH_exp_data_plot}. The tape presents a quite regular dependence on the orientation of the magnetic field, with the maxima of $I_c$ obtained with a field almost parallel to the tape, and the minima with a field perpendicular to the tape. On the other hand,  the reduction of the critical current with the field angle does not follow the natural symmetry of the tapes, and the $I_c$ values are slightly different in the four quadrants of the angle span. However, since the primary purpose of this work is to show the principles of use of a method for extracting the angular dependence of the critical current density from measured $I_c$ values, we chose a simpler set of experimental data by using a symmetric angular dependence. This was done by averaging the data of the four quadrants -- see data points in Fig.~\ref{fig:best7} later in the paper.

Our method consists in the following steps.

\begin{enumerate}
\item Choose a functional $J_c(B,\theta)$ dependence, based on the shape of the experimental data set;
\item Choose a numerical model that, with the chosen $J_c(B,\theta)$, can compute the effective critical current of the tape for a given field amplitude and orientation;
\item Tune the parameters (either manually or automatically) until the calculated $I_c(H_a,\theta)$ data sets matches the experimental ones reasonably well.
\end{enumerate}

A good solution consists in using a field-dependent critical state model, either home-made, such as in \cite{Rostila:PhysC07}, or implemented in a commercial finite element program, such as in \cite{Gomory:SST06b} (Magnetostatic module of {\it FEMLAB} -- now become {\it COMSOL} -- in that case). The model solves in 2-D the equation for the magnetic vector potential $\mathbf{A}$
\begin{equation}
\nabla \times \left ( \frac{1}{\mu} \nabla \times \mathbf{A} \right )=\mathbf{J}
\end{equation}
and is based on an iterative resolution of the $J$ distribution with the constraint that $J=J_c(\mathbf B)$, and where the spatial distribution of ${\mathbf B}$ in the tape cross-section ($x-y$ plane) is refreshed at each iteration, until self-consistent $J_c({\mathbf B(x,y)})$ and ${\mathbf B}(x,y)$ distributions are achieved. The external magnetic field of the desired magnitude and orientation is applied by appropriately setting the value of the magnetic vector potential on the domain's boundary.

\begin{table}[t!]
\renewcommand\arraystretch{1.3}
\caption{\label{tab:param}Tested parameters.}
\centering
\begin{tabular}{lll}
Parameter &  Tested values &  \# of values \\ \hline
$J_{c0}$ (A/m$^2$)	& [4, 4.25, 4.5, 4.75. 5, 5.25]  $\cdot10^{10}$~~ 	& 6 \\
$B_c$ (mT)		& 20, 25, 30, 35, 40 						& 5 \\
$b$				& 0.3, 0.4, 0.5, 0.6, 0.7, 0.8, 0.9				& 7 \\
$k$				& 0.05, 0.1, 0.15, 0.2, 0.25, 0.3				& 6 
\end{tabular}
\end{table}

\begin{table}[t!]
\renewcommand\arraystretch{1.3}
\caption{\label{tab:best7} Best seven sets of parameters.}
\centering
\begin{tabular*}{\columnwidth}{llllllll}
Set \# &$J_{c0}$  	& $B_c$ & $b$ & $k$ & $\epsilon$ & $\epsilon_{avg}$ &$I_c$  \\ 
 & (A/m$^2$) & (mT) & & & & (\%) & (A) \\
\hline \\
1	& 4.75$\cdot10^{10}$ 	& 35	& 0.6	 & 0.25	& 8.41E-04	& 1.95 & 158.2\\
2	& 5.00$\cdot10^{10}$ 	& 30	& 0.6	 & 0.25	& 8.57E-04	& 2.33 &  161.4\\
3	& 5.00$\cdot10^{10}$ 	& 40	& 0.7	 & 0.30	& 9.74E-04	& 2.52 & 163.4\\
4	& 4.75$\cdot10^{10}$ 	& 35	& 0.6	 & 0.30	& 1.13E-03	& 2.96 & 157.7\\
5	& 4.50$\cdot10^{10}$ 	& 40	& 0.6	 & 0.25	& 1.16E-03	& 2.79 & 153.6\\
6	& 5.00$\cdot10^{10}$ 	& 30	& 0.6	 & 0.30	& 1.23E-03	& 2.95 & 160.9\\
7	& 5.25$\cdot10^{10}$ 	& 20	& 0.5	 & 0.30	& 1.25E-03	& 2.81 & 162.9
\end{tabular*}
\end{table}

We tested our method on coated conductor tapes exhibiting a the angular dependence shown in Fig.~\ref{fig:best7}, for which one can assume an elliptical functional form of $J_c(B,\theta)$: 
\begin{equation}\label{eq:elliptical}
J_c(B_\parallel,B_\perp)=\frac{J_{c0}}{\left [1+\sqrt{(k B_\parallel)^2+B_\perp^2}\,/B_c\right ]^b} \,,
\end{equation}
where $J_{c_0}$, $k$, $B_c$ and $b$ are the four parameters used here for the curve fitting.
In this case, instead of tuning the parameters manually, we used an automatic method to find the optimal parameters sets, as follows.
We chose a range of variation for the parameters in the $J_c(B,\theta)$ relation and ran the model according to a table that generates all possible combinations of the parameters and of field amplitude and orientation (these latter are the same as the experimental ones). Then, for each simulated case, we calculated the error with respect to the corresponding experimental data point (one data point $=$ one value of field and angle); finally, we calculated the average error associated with each simulated parameter set of the $J_c(B,\theta)$, so that the best parameter sets can be determined.

The ranges of values chosen for sweeping the parametric space with these parameters are summarized in Table~\ref{tab:param}. The total number of combinations is \hbox{$6 \!\times\! 5 \!\times\! 7 \!\times\! 6=1260$}. For each combination of parameters, simulations were performed for 14 different angles of the magnetic field and 3 magnitudes (as many as the experimental data points after symmetrization, see Fig.~\ref{fig:best7}). The total number of simulated cases was then \hbox{$14 \!\times\! 3 \!\times\! 1260=52920$}. This may seem a large number, but the  model runs in 1 second or less per case, so all cases could be simulated in a few hours by putting the computations in parallel on a standard desktop workstation.

In order to find the best set of parameters, we computed the mean quadratic error with respect to the experimental data points

\begin{equation}\label{eq:error}
	\epsilon=\frac{1}{N_{H_a}N_{\theta}}\sum_{N_{H_a}N_{\theta}} \,
	\left [ \frac{I_{c_{calc}}-I_{c_{meas}}}{I_{c_{meas}}}\right ]^2,
\end{equation}
where $N_{H_a}$ and $N_{\theta}$ are the total numbers of field amplitudes and orientations, respectively (in our case \hbox{$N_{H_a}=3$} and \hbox{$N_{\theta}=14$}), and where $I_{c_{calc}}$ and $I_{c_{meas}}$ are the calculated and measured critical currents, respectively.

\begin{figure}[t!]
\centering
\includegraphics[width=8 cm ]{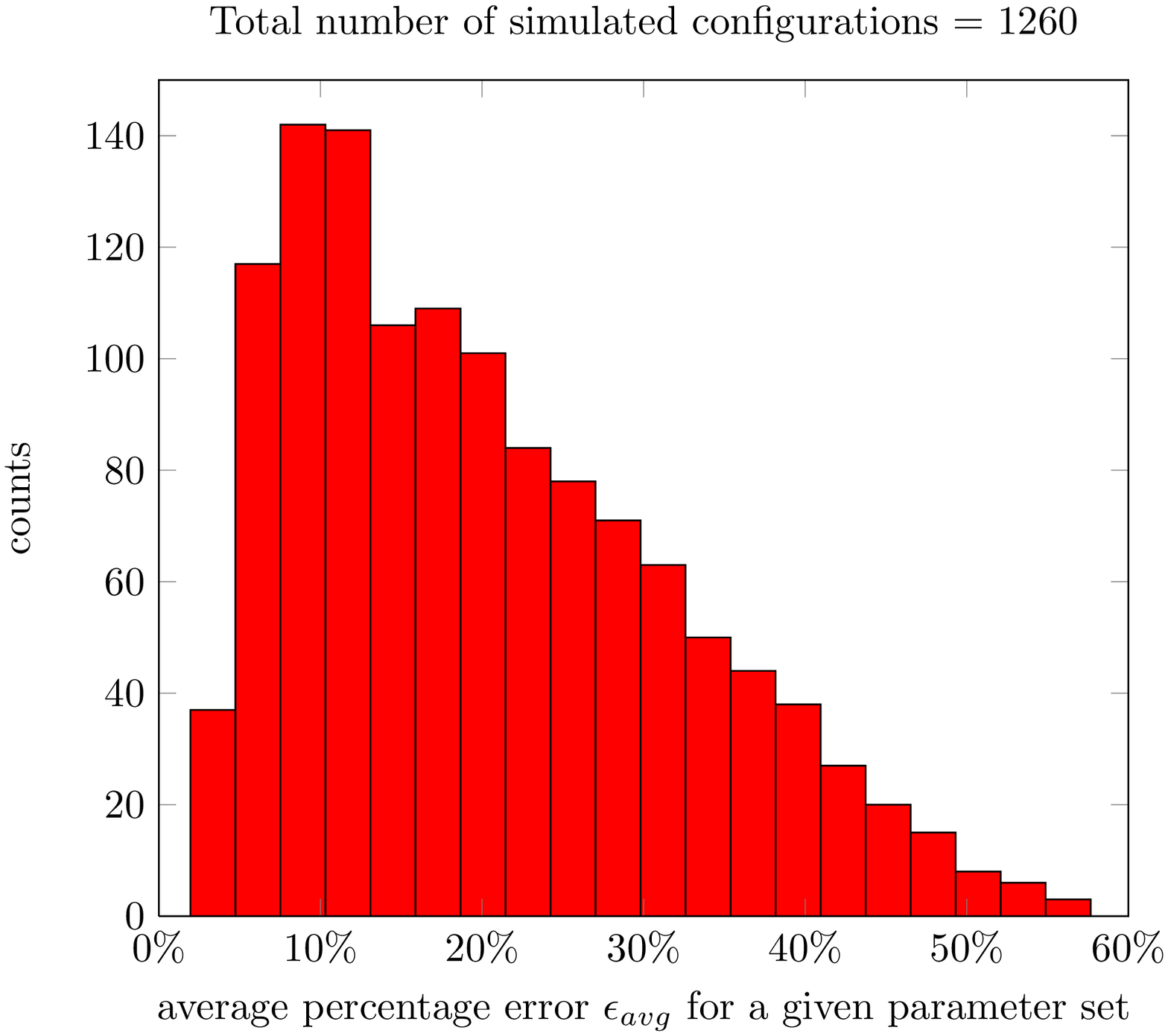}
\caption{\label{fig:histogram}{Statistical distribution of the average percentage of error for all simulated cases with respect to the experimental data.}}
\end{figure}

There are several sets of parameters that give an equally good match of the experimental data. Table~\ref{tab:best7} lists the best seven sets of parameters. In addition to the mean quadratic error, the table displays the average error $\epsilon_{avg}$ with respect to the experimental data points. This latter is computed as 

\begin{equation}\label{eq:error_avg}
	\epsilon_{avg}=\frac{1}{N_{H_a}N_{\theta}}\sum_{N_{H_a}N_{\theta}} \,
	\left | \frac{I_{c_{calc}}-I_{c_{meas}}}{I_{c_{meas}}}\right |.
\end{equation}
As shown in the table, the best seven parameter sets have a very low average error, below 3~\%, and the calculated angular dependence of the critical current are not very different and reproduce the experimental data point reasonably well -- see Fig.~\ref{fig:best7}. 
Indeed, there are many parameters sets that give a low average error. This is shown in Fig.~\ref{fig:histogram}, which shows the statistical distribution of $\epsilon_{avg}$ for all 1260 simulated cases. One can see that there are more than 200 parameter sets with an average error of less than 10~\%, which is an indication that the choice of the most precise $J_c(B,\theta)$ is not so important. In that respect, a manual tuning of the parameters is equally effective. One has also to remember that tapes (especially long ones as those employed in superconducting devices) exhibit a statistical variation of the critical current along their length of a few percent. So any choice of parameters in the $J_c(B,\theta)$  able to provide this kind of accuracy can be considered satisfactory. 

\section{Computation of effective $I_c$ of superconducting devices }\label{sec:effIc}

Once the form of the $J_c(B,\theta)$ material model is selected and its parameters are determined with the procedure described in the previous section, it can be used as input to calculate the effective critical current of devices, which is determined by the electromagnetic interaction of the many tapes composing them. 
In this section we consider three cable samples: a Roebel cable composed of strands punched from the same tapes as those considered in the previous section; a cable made of 6 mm-wide strands with higher current-carrying capacity, characterized by a more complicated angular dependence of the critical current of the composing strands; a laboratory-scale single-phase cable composed of Bi-2223 tapes.

In applying the model to a geometry such that of a Roebel cable, we assume that the cable reaches its critical current when its complete cross-section is filled with the critical current density. This represents an optimistic scenario, based on the assumption that the current can distribute between the strands. Other scenarios, like the complete electrical insulation between the strands, can be considered by using specifically developed approaches~\cite{Vojenciak:SST11}. 
These latter, however, fall behind the scope of this work, which is to test the accuracy of the current model for computing the critical current in cables and to evaluate the importance of using a very accurate description of the angular dependence of the critical currents of HTS tapes).

An important factor that can influence the current repartition in real cables is the contact resistance. Earlier studies revealed the importance of using sufficiently long soldered current contacts to make sure that the current repartition is balanced~\cite{Goldacker:SST09}. In the Roebel cables presented here the length of the current contact was more than one transposition length. In the case of the Bi-2223 power cable, equal current distribution between the strands is guaranteed by a series connection, as mentioned later. 

\subsection{Low-current Roebel cable} 
The cable is composed of 10 strands, punched from 4~mm-wide tapes. This kind of cable is for example used in Rutherford-like cable prototype for high current applications~\cite{Kario:SST13}. The cable is approximated with its 2-D cross-section, i.e. as two stacks of five strands each, as illustrated in Fig.~\ref{fig:drawing}. Each strand is 1.8 mm wide and 1 $\mu$m thick. The central gap is 0.4~mm and the vertical separation between the superconducting layers is 100~$\mu$m. For the superconductor's properties we used the parameters of set \#1 in Table~\ref{tab:best7}. 

\begin{figure}[t!]
\centering
\includegraphics[width= 8 cm ]{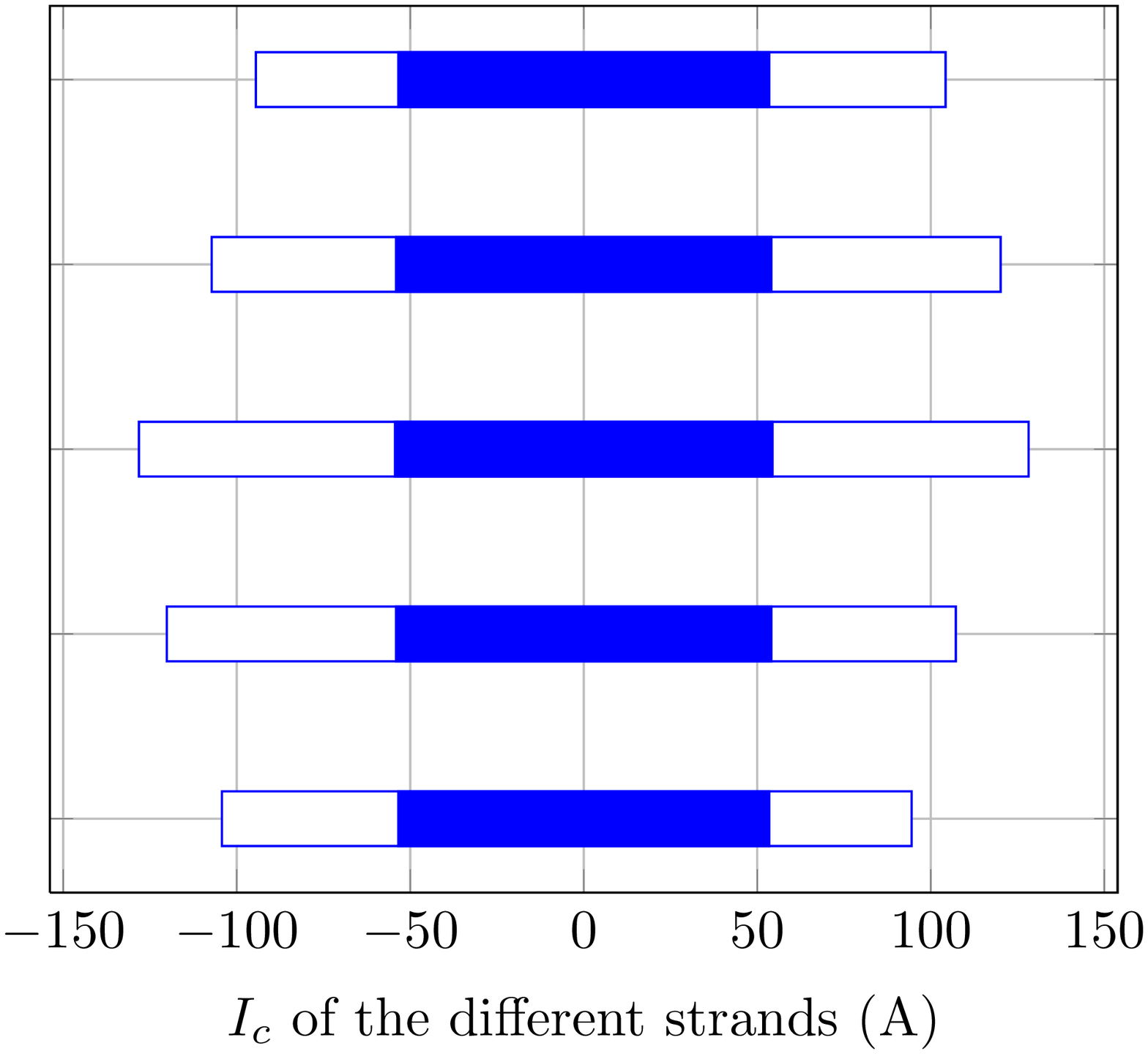}
\caption{\label{fig:strands_Ic_plot}{Calculated critical current in the different strand positions for the low-current (blue bars) and high-current (white bars) Roebel cables. The low-current cable was simulated with the parameter set \#1 of Table~\ref{tab:best7}. The high-current cable with the set given in the upper part of Table~\ref{tab:IcHplot_HighCurrent}.}}
\end{figure}

With the assumptions of this model, every point in the superconductor is at its critical current density and the effective critical current is given by integrating the distribution of $J_c({\bf B}(x,y))$ over the superconductor's cross section.
With the utilized elliptical $J_c(B,\theta)$, the integral of the critical current density in the different strand positions, gives very similar results, as shown by the solid bars in Fig.~\ref{fig:strands_Ic_plot}.
The computed critical current for the whole cable is 539 A. This value is much lower than the critical current that is obtained by simply adding the critical current of each strand, i.e. about 710 A, which indicates the important influence of the self-field in reducing the current-carrying capability of the cable. This value was confirmed by simulations carried out with a transient model using power-law resistivity for the description of the superconductor material -- see appendix for details.

\begin{figure}[t!]
\centering
\includegraphics[width=8 cm ]{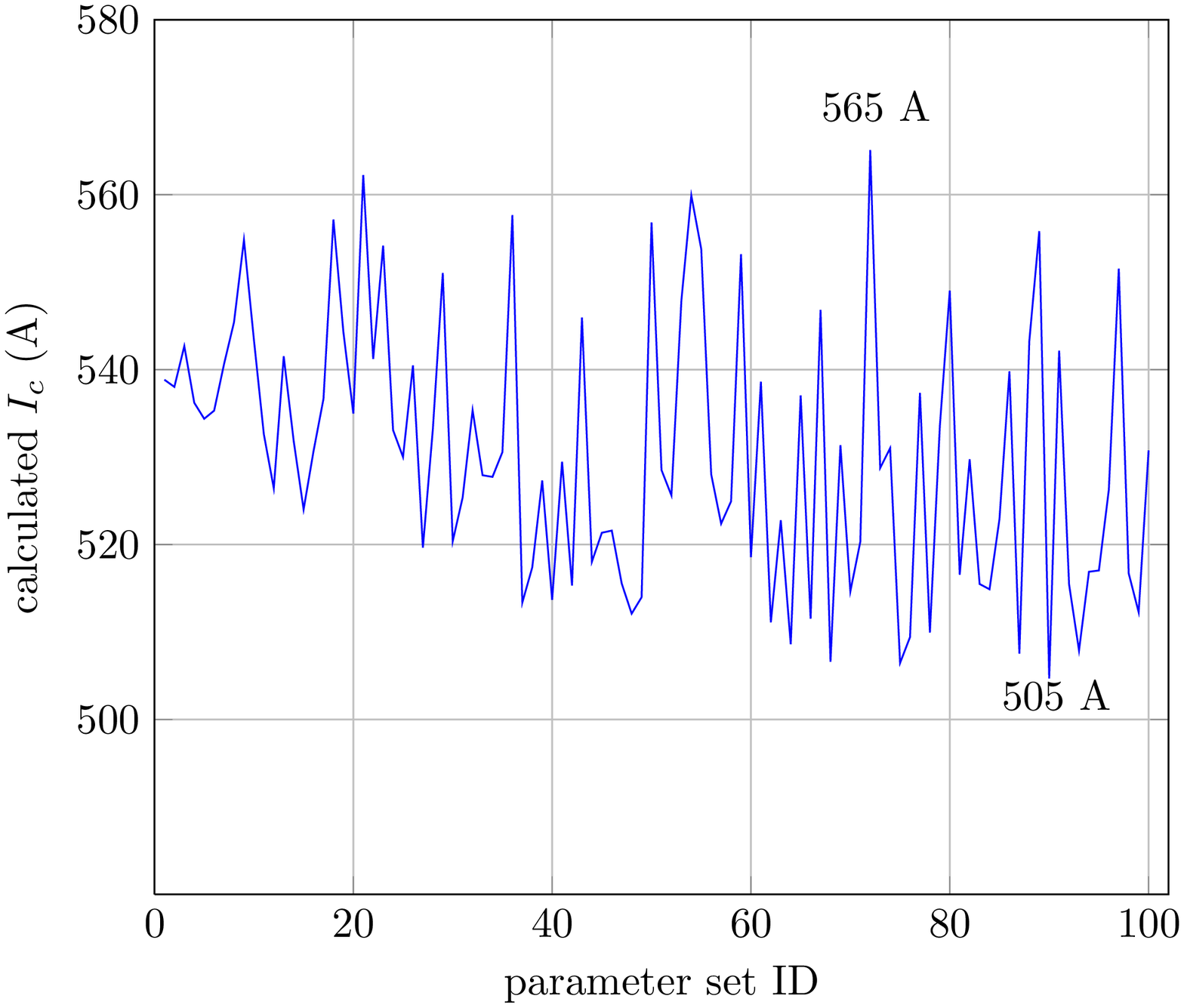}
\caption{\label{fig:Ic_Roebel_best100}{Critical current of the Roebel cable calculated for the best 100 parameter sets of the $J_c(B,\theta)$ of a single tape: the highest and lowest value are 565 A and 505 A, respectively.}}
\end{figure}

The fact that, as shown at the end of section~\ref{sec:extraction}, there are many parameter sets giving a very similar small average deviation from experimental value of $I_c(H_a,\theta)$ has the consequence that also the critical current of devices computed using those parameter sets does not change much. An example is shown in Fig.~\ref{fig:Ic_Roebel_best100}, which shows the calculated critical currents of the Roebel cable for the 100 best parameters sets of the $J_c(B,\theta)$ of a single tape. With this relatively wide choice of parameter sets, the calculated critical currents fall within 36 A from the value calculated with set \#1, which corresponds only to a 7~\% span.

It is  worth noting that using a $J_c(B,\theta)$ functional dependence able to closely reproduce the asymmetrical measured angular dependence of $I_c$ (namely, the fact the the peaks of $I_c$ are not equally high and occur at slightly different angles than 0$^{\circ}$ and 180$^{\circ}$) does not have  practical impacts:
other than amplifying the difference between the critical currents of the different strand positions~\cite{Kario:SST13}, the use of a more precise (and asymmetrical) $J_c(B,\theta)$ predicts a  critical current of the cable a few amperes different from that calculated here (542 A against 539 A), which is similarly distant from the experimentally measured value of 460 A.

This latter discrepancy between calculated and measured $I_c$ has to be put in relation with tape uniformity and current degradation due to the punching process. As reported in~\cite{Kario:SST13}, the measured critical current of the strands composing the cable exhibit an important scattering of values, ranging from 54.3 to 71 A with an average value of 61.4 A. If simulations for the cable are repeated by scaling $J_{c0}$ by a factor 61.4/71=0.865 to take into account this reduction of the strands'  critical current, the estimated critical current of the cable goes down to 483~A, which is within 5~\% of the measured value. 

\subsection{High-current Roebel cable}
\begin{figure}[t!]
\centering
\includegraphics[width=8 cm ]{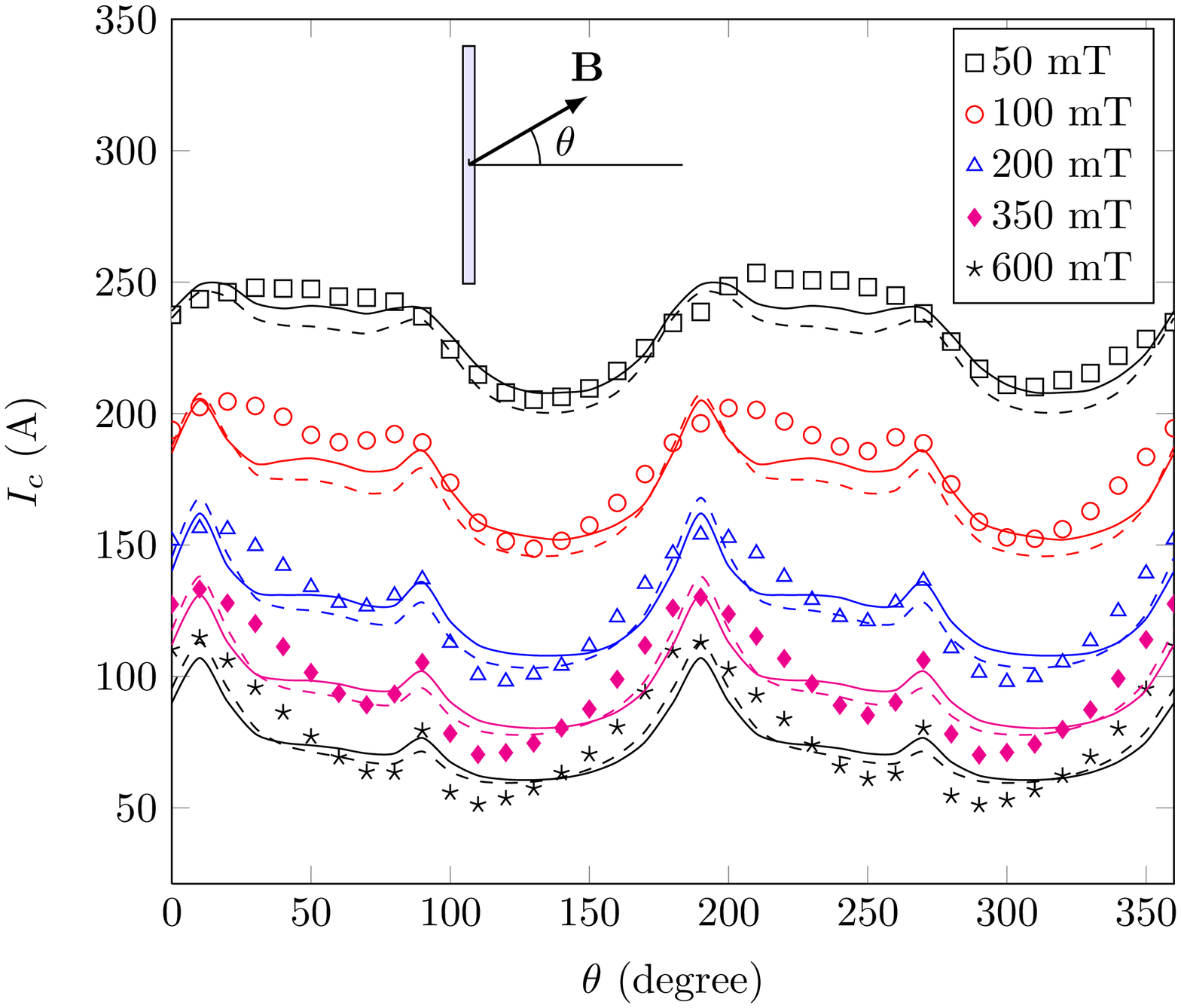}
\caption{\label{fig:IcHplot_HighCurrent}{Comparison of experimental (symbols) and calculated (lines) angular dependence on the applied magnetic field for the coated conductors used to assemble the high-current Roebel cables. The continuous and dashed lines represent the calculated values with the two set of parameters in the $J_c(B,\theta)$ dependence given in Table~\ref{tab:IcHplot_HighCurrent}. The self-field critical currents calculated with those two sets are 340 and 339 A, respectively.}}
\end{figure}

This Roebel cable is composed of 10 strands, punched from 12~mm-wide tapes, which have a self-field critical current of 340 A and exhibit a more complicated angular dependence~\cite{Grilli:TAS14b}.  Also in this case different sets of parameters can be chosen to give an equally good match of the experimentally measured $J_c(B,\theta)$ for the single tape, see Fig.~\ref{fig:IcHplot_HighCurrent}.
Table~\ref{tab:IcHplot_HighCurrent} lists the two sets of parameters used in the calculated angular dependence of Fig.~\ref{fig:IcHplot_HighCurrent}. For the analytical formula of $J_c(B, \theta)$, see~\cite{Grilli:TAS14b} (table II there has a missing column, which has been correctly inserted here)  . The average error with respect to the experimental data (averaged over all the 18$\times$5=180 data points) is 6.8~\% and 6.7~\% for the continuous and dashed curve, respectively.

\begin{table}[ht!]
\renewcommand\arraystretch{1.3}
\caption{\label{tab:IcHplot_HighCurrent}Coefficients of the $J_c(B, \theta)$ dependence for tape of Fig.~\ref{fig:IcHplot_HighCurrent}.}
\centering
\begin{tabular}{llllll}
$x$ 	& $J_{c0_{x}}~ {\rm (A/m^2)}$	& $B_{0x}$ (T) & $\beta_{x}$	&  $d_x$ (rad)	& $u_x$ \\ \hline \hline
\multicolumn{6}{c}{Continuous line (from~\cite{Grilli:TAS14b})}\\
90	& 3.2 $\cdot 10^{10} $ & 0.035	  & 0.57 	& 1.5708  & 6  \\ \hline 
10	& 3.5 $\cdot 10^{10} $ & 0.020	  & 0.4	& 0.1745  & 7  \\ \hline 
50	& 3.5 $\cdot 10^{10} $ & 0.032    & 0.59 	& 0.8727  & 1.5  \\ \hline \hline 
\multicolumn{6}{c}{Dashed line}\\
90	& 4.0 $\cdot 10^{10} $ & 0.020	& 0.57	& 1.5708  & 6  \\ \hline 
10	& 3.0 $\cdot 10^{10} $ & 0.035	& 0.4 	& 0.1745  & 7  \\ \hline 
50	& 3.6 $\cdot 10^{10} $ & 0.027	& 0.59	& 0.8727  & 1.5  \\ \hline \hline 
\end{tabular}
\end{table}

The self-field critical current of the Roebel cable at 77~K is 1002 A. This value is higher than that reported in~\cite{Grilli:TAS14b} (936 A), because in that case the critical current was measured using a lower electric field criterion (to avoid possible damage of the cable). The value of 1002 A refers to the 1~$\mu$V/cm criterion, which is the same criterion used for the $J_c(B,\theta)$ characterization of the composing tapes. In this case the parameters of the $J_c(B,\theta)$ were determined by manual tuning.

The calculated critical currents with the parameters sets of Table~\ref{tab:IcHplot_HighCurrent} are 1109  and 1087 A, respectively, which are only 11 and 8 \% away from the measured value.

This discrepancy can be considered acceptable, given the number of possible causes that can concur to modify the critical current value and are not taken into account by our model: in addition to the already mentioned absence of current sharing between the strands, they include for example the possible slight degradation of the transport properties of the composing strands due to the punching process (in this case the individual strands were not characterized individually) and the tape misalignment in the cable structure.

For this cable the integral of $J_c(B,\theta)$ in the different strand positions gives quite different values, as shown by the white bars in Fig.~\ref{fig:strands_Ic_plot}. In particular, for the two sets of parameters  listed in Table~\ref{tab:IcHplot_HighCurrent}, those values range from 94 to 128 A and from 92 to 119 A, respectively. It is interesting to note the asymmetric distribution of those values, which is a direct consequence of the angular asymmetry of the $J_c(B,\theta)$ utilized in this case.

\subsection{Bi-2223 power cable}
The cable is a laboratory-scale model of the cable used in the Ampacity project~\cite{Ampacity_website, Elschner:EUCAS13}. The $I_c$ characterization was performed on a single phase, where the 22 Bi-2223 tapes are connected in series (by means of return copper conductors) to make sure that the same current flows in all the tapes and the current repartition is not unbalanced due to the influence of the contact resistance, as it is common on short cable samples. The polygonal arrangement of the tapes around a central cylindrical former is known to have a positive effect on the transport properties of the tapes due to the partial cancellation of the field component perpendicular to the tapes \cite{Siahrang:TAS10}.

\begin{figure}[t!]
\centering
\includegraphics[width=8 cm ]{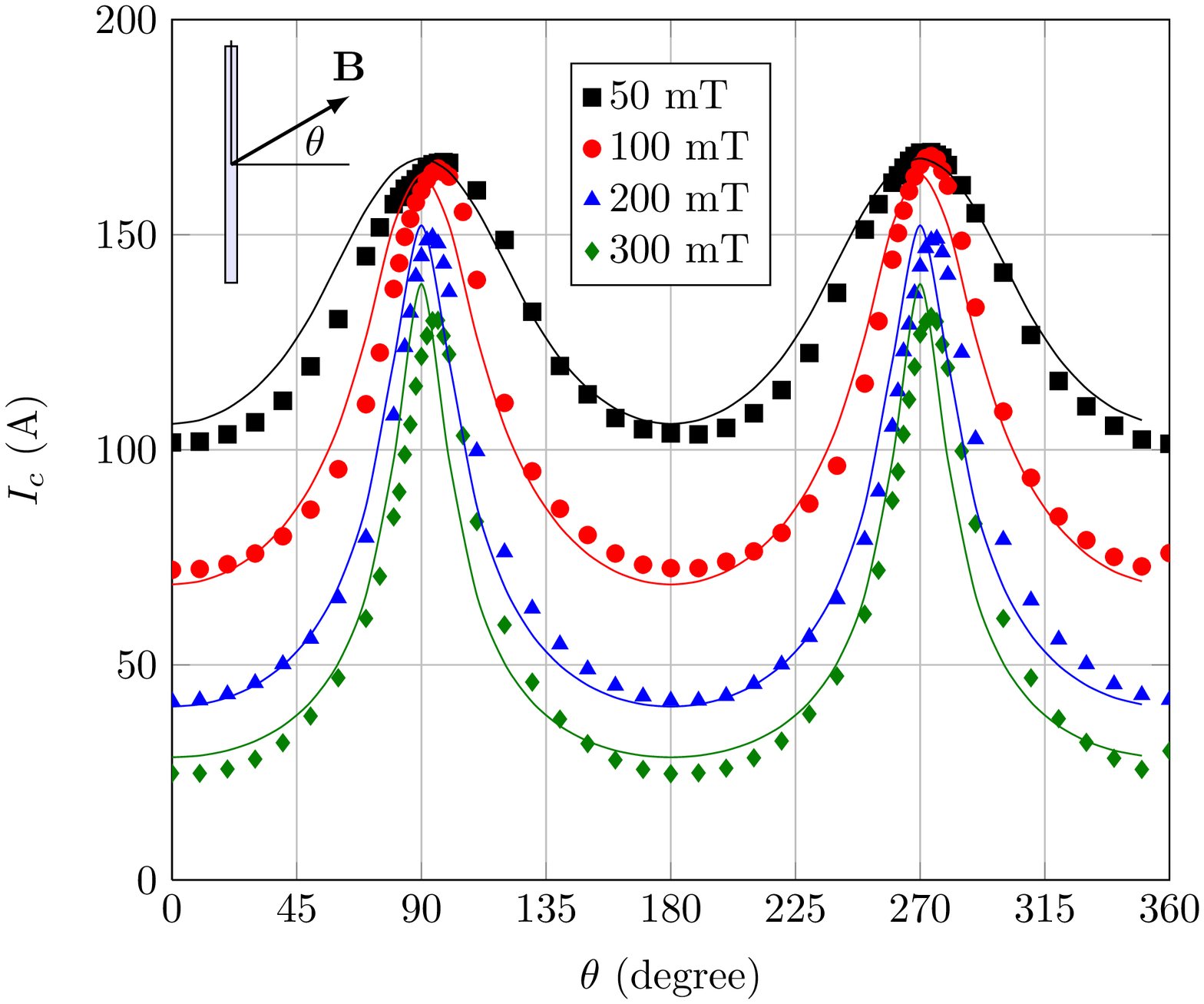}
\caption{\label{fig:BSCCO_IcH_plot}{Comparison of experimental (symbols) and calculated (continuous lines) angular dependence on the applied magnetic field for the Bi-2223 tapes used in the power cable sample.}}
\end{figure}

The tapes exhibit an elliptical dependence of the critical current on the magnetic field, which can be described by~\eqref{eq:elliptical} with the following parameters: \hbox{$J_{c0}=3.7\cdot10^{8}$~(A/m$^2$)}, $B_c=42$~mT, $k=0.13$
, $b=1$. The comparison of the calculated and measured $I_c$ at a given amplitude and field is shown in Fig.~\ref{fig:BSCCO_IcH_plot}. For the self-field critical current of the isolated tapes an average value of 167$\pm$2 was found. 
The corresponding calculated critical current is 169 A.

The tapes's critical current was successively measured with all the tapes energized (and carrying the same current) and resulted to be 180$\pm$3. The computed value is 179 A. Again, the proposed model showed its effectiveness for computing the effective critical current of cables composed of interacting tapes.

\section{Conclusion}
In this paper we presented a simple method to extract the $J_c(B,\theta)$ dependence from experimental characteristics of the angular dependence of the critical current as a function of the applied field and to calculate the effective critical current of superconducting devices composed of electromagnetically interacting tapes.

The method consists in using a fast numerical model to find the parameters of the $J_c(B,\theta)$ functional dependence of the superconducting material. The model, based on the critical state with field-dependent $J_c$, for a given parameter set and applied magnetic field, calculates the effective critical current of the tape, which includes the self-field effects.
We found that there are many different sets of parameters that give a critical current for the tape very close to the experimentally measured value, with an average distance from the experimental data points of only a few percent. What is important to note is that, in these sets giving good results, the parameter values can vary quite considerably. 

Successively we employed the same numerical method to calculate the effective critical current of three different cables: two differently current rated Roebel cables and a laboratory-scale prototype of a power transmission cable. The model proved to be very efficient in calculating the critical current of the cables, which resulted to be only slightly higher (maximum error 11~\%) than the measured value.

The results obtained in this work indicate that the determination of a very precise set of parameters for angular dependence $J_c(B,\theta)$ is not so important, as long as the main features of the dependence are reproduced. In this respect, one has also to keep in mind that various effects in real samples can importantly influence the actual value of the critical current and reduce the importance of a very precise $J_c(B,\theta)$ model: as seen in the experimental examples presented in this work, these include uniformity of the tape properties, degradation due to the assembling process, imprecisions in the positioning of the tapes in the superconducting device.

The work has been carried out with {\it RE}BCO coated conductors and Bi-2223 tapes, however the method developed for the extraction of the $J_c(B,\theta)$ characteristic from experimental data and the computation of effective $I_c$ in tape assemblies can be employed with any type of superconductor. The easiness of use of the model -- which can be easily implemented in different software packages -- and its high computing speed make it a very attractive tool for the purpose of characterizing superconducting tapes and calculate the effective critical current of superconducting devices.

\section*{Appendix}
In order to validate our results not only against experiments, but also against another model, we performed the critical current calculations with a time-dependent FEM model based on the $H$-formulation of Maxwell equations~\cite{Brambilla:SST07}. The model is usually utilized to solve transient problems (in particular to compute  AC losses), but different authors have used it also to calculate the effective critical current of tape assemblies~\cite{Thakur:SST11b, Zhang:JAP12}.  
Since it solves transient problems, it can  be used for reproducing DC $V-I$ curves by applying a current ramp from zero to the final value, and then let the current distribution relax to an apparent steady-state, i.e. a state in which further current relaxation is unnoticeable on the time scale of interest, typically a few seconds. Since the variable solved for is the magnetic field $H$, the model easily takes into account the field dependence of $J_c$.
Being time-dependent, the simulations carried out with this model are typically very long (up to several hours in the case of many interacting tapes). Therefore the model is used here just to validate some of the results computed with the fast model presented in this paper.

\begin{figure}[t!]
\centering
\includegraphics[width= 8 cm ]{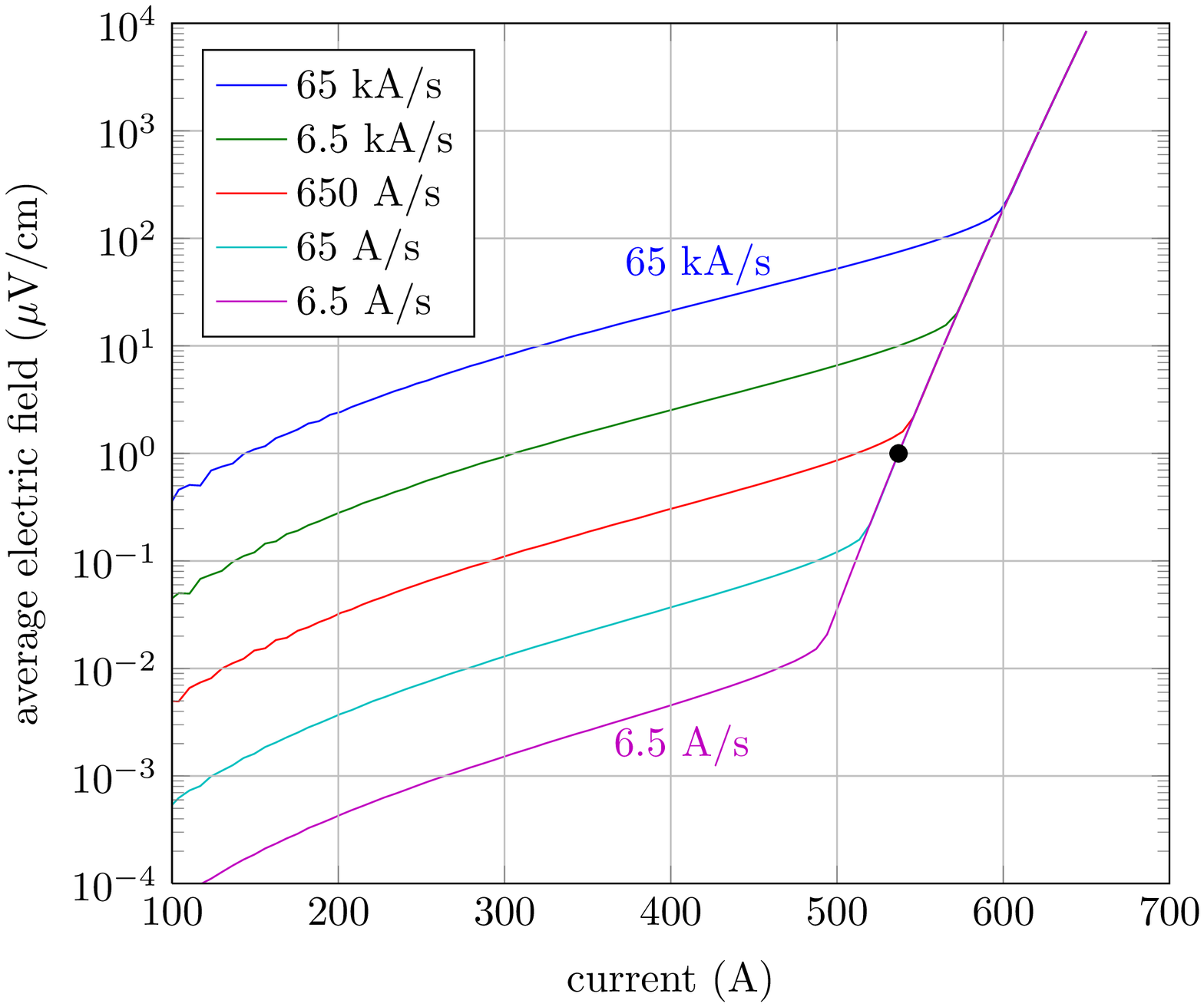}
\caption{\label{fig:HK_ramp}{$E-I$ characteristics for a Roebel cable calculated with the transient $H$-formulation model using current ramps of different speeds. Only the slowest ramps give the correct value of the critical current (537 A, indicated by the black dot), taken when the average field reaches $1~\mu$V/cm.}}
\end{figure}

In the $H$-formulation FEM model, we applied a current ramp and measured the average electric field in the superconductor, taking the current at a critical average field $E_0=1~\mu$V/cm as the critical one. Due to the relaxation of the magnetic field inside the superconductor, the ramp must be sufficiently slow to simulate a DC situation. 
An example is shown in Fig.~\ref{fig:HK_ramp}, which shows the calculated $E-I$ characteristics of the low-current Roebel cable for different speeds of current ramp from 0 to $I_{max}=650$~A: only the slowest ramps converge to the correct value of the critical current. The obtained value of 537 A, indicated with a black dot in the figure, is very close to the 539 A obtained with the static method. Similar results for the shape of the $E-I$ curves have been reported by Thakur {\it et al.}~\cite{Thakur:SST11b}.

\begin{figure}[h!]
\centering
\includegraphics[width=8 cm ]{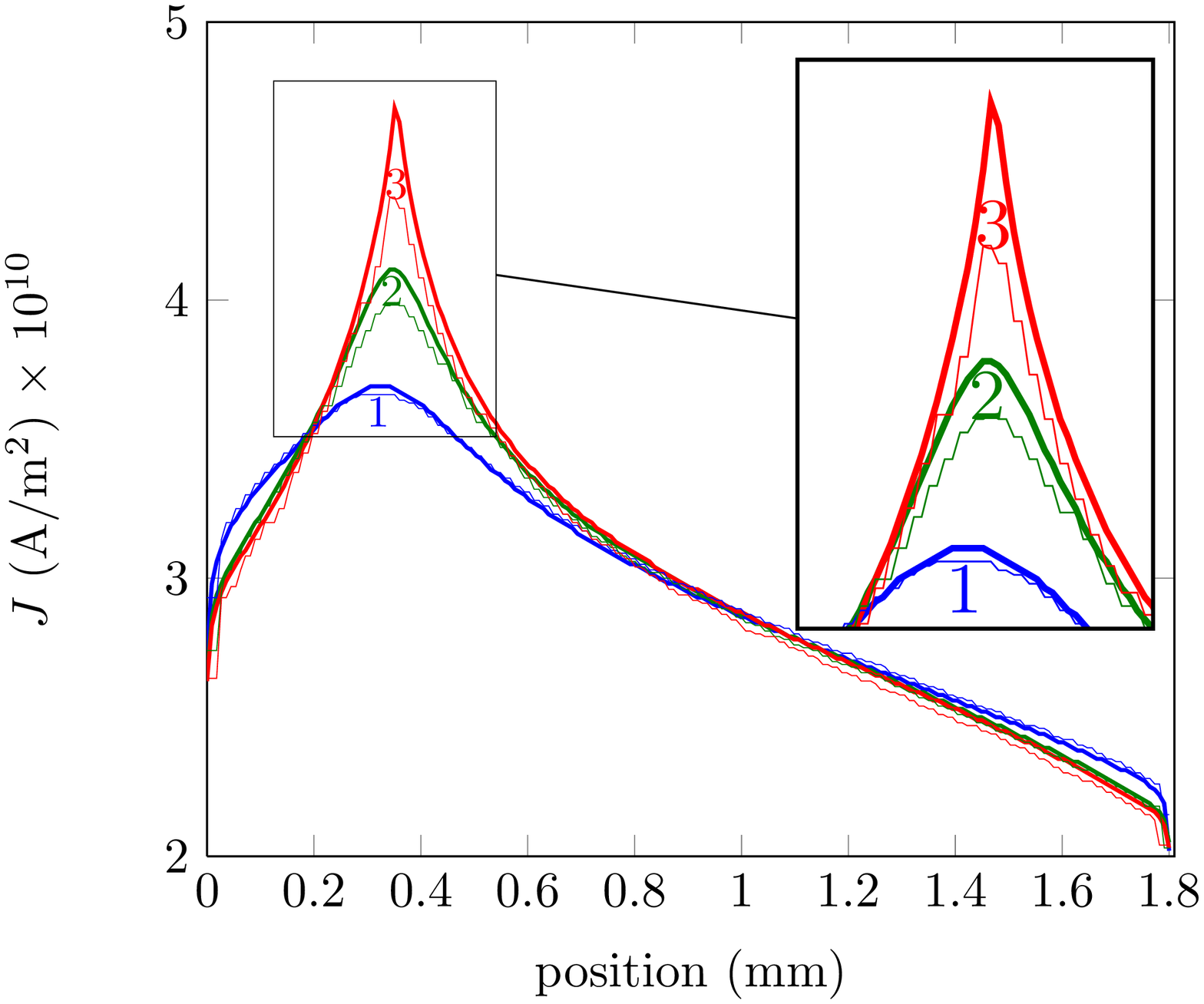}
\caption{\label{fig:J_profiles_plot}{Comparison of the current density distribution at $I=I_c$ computed with the static critical state (thick lines) and transient $H$-formulation  (thin lines) models. The profiles are taken along the width of tapes number 1, 2, 3 of Fig.~\ref{fig:drawing}. }}
\end{figure}

In addition to the values of the critical current, one can compare the two models also on local quantities, such as current density and magnetic field distributions. Fig.~\ref{fig:J_profiles_plot} shows the local values of $J$ in the two models. The profiles are taken along the width of tapes number 1, 2, 3 of Fig.~\ref{fig:drawing}. The results of the two models are in good agreement, the profile computed with the CSM being a little sharper, due to the use of the critical state model. One can observe that the profile computed with the transient $H$-formulation model is a little jagged. This is due to the model's feature of computing piecewise constant values of $J$ in each element of the mesh -- see~\cite{Zermeno:JAP13} for more details.
%



\begin{thebibliography}{10}
\providecommand{\url}[1]{#1}
\csname url@samestyle\endcsname
\providecommand{\newblock}{\relax}
\providecommand{\bibinfo}[2]{#2}
\providecommand{\BIBentrySTDinterwordspacing}{\spaceskip=0pt\relax}
\providecommand{\BIBentryALTinterwordstretchfactor}{4}
\providecommand{\BIBentryALTinterwordspacing}{\spaceskip=\fontdimen2\font plus
\BIBentryALTinterwordstretchfactor\fontdimen3\font minus
  \fontdimen4\font\relax}
\providecommand{\BIBforeignlanguage}[2]{{%
\expandafter\ifx\csname l@#1\endcsname\relax
\typeout{** WARNING: IEEEtran.bst: No hyphenation pattern has been}%
\typeout{** loaded for the language `#1'. Using the pattern for}%
\typeout{** the default language instead.}%
\else
\language=\csname l@#1\endcsname
\fi
#2}}
\providecommand{\BIBdecl}{\relax}
\BIBdecl

\bibitem{MacManus-Driscoll:NMAT04}
J.~L. MacManus-Driscoll, S.~R. Foltyn, Q.~X. Jia, H.~Wang, A.~Serquis,
  L.~Civale, B.~Maiorov, M.~E. Hawley, M.~P. Maley, and D.~E. Peterson,
  ``{Strongly enhanced current densities in superconducting coated conductors
  of YBa2Cu3O7-x + BaZrO3.}'' \emph{Nature materials}, vol.~3, no.~7, pp.
  439--443, 2004.

\bibitem{Maiorov:APL05}
B.~Maiorov, B.~J. Gibbons, S.~Kreiskott, V.~Matias, T.~G. Holesinger, and
  L.~Civale, ``{Effect of the misalignment between the applied and internal
  magnetic fields on the critical currents of ``tilted coated conductors''},''
  \emph{Applied Physics Letters}, vol.~86, no.~13, p. 132504, 2005.

\bibitem{Holesinger:SST09}
T.~G. Holesinger, B.~Maiorov, O.~Ugurlu, L.~Civale, Y.~Chen, X.~Xiong, Y.~Xie,
  and V.~Selvamanickam, ``{Microstructural and superconducting properties of
  high current metal--organic chemical vapor deposition $\rm
  YBa_2Cu_3O_{7−\delta}$ coated conductor wires},'' \emph{Superconductor
  Science and Technology}, vol.~22, no.~4, p. 045025, 2009.

\bibitem{Zhang:PhysC09}
Y.~Zhang, E.~D. Specht, C.~Cantoni, D.~K. Christen, J.~R. Thompson, J.~W.
  Sinclair, A.~Goyal, Y.~L. Zuev, T.~Aytug, M.~P. Paranthaman, Y.~Chen, and
  V.~Selvamanickam, ``{Magnetic field orientation dependence of flux pinning in
  $\rm (Gd,Y)Ba_2Cu_3O_{7−x}$ coated conductor with tilted lattice and
  nanostructures},'' \emph{Physica C}, vol. 469, no. 23-24, pp. 2044--2051,
  2009.

\bibitem{Selvamanickam:PhysC09}
V.~Selvamanickam, Y.~Chen, J.~Xie, Y.~Zhang, A.~Guevara, I.~Kesgin, G.~Majkic,
  and M.~Martchevsky, ``{Influence of Zr and Ce doping on electromagnetic
  properties of (Gd,Y)--Ba--Cu--O superconducting tapes fabricated by metal
  organic chemical vapor deposition},'' \emph{Physica C: Superconductivity},
  vol. 469, no. 23-24, pp. 2037--2043, 2009.

\bibitem{Selvamanickam:SST12}
V.~Selvamanickam, Y.~Yao, Y.~Chen, T.~Shi, Y.~Liu, N.~D. Khatri, J.~Liu,
  C.~Lei, E.~Galstyan, and G.~Majkic, ``{The low-temperature,
  high-magnetic-field critical current characteristics of Zr-added $\rm
  (Gd,Y)Ba_2Cu_3O_x$ superconducting tapes},'' \emph{Superconductor Science and
  Technology}, vol.~25, no.~12, p. 125013, 2012.

\bibitem{Rostila:SST07}
L.~Rostila, J.~Lehtonen, R.~Mikkonen, J.~{\v S}ouc, E.~Seiler, T.~Mel{\'\i}{\v
  s}ek, and M.~{Vojen{\v c}iak}, ``How to determine critical current density in
  {YBCO} tapes from voltage--current measurements at low magnetic fields,''
  \emph{Superconductor Science and Technology}, vol.~20, pp. 1097--1100, 2007.

\bibitem{Sirois:ACE02}
F.~Sirois, D.~R. Watson, W.~Zhu, and J.~R. Cave, ``Development of a numerical
  method to determine the local {E-J} characteristics of anisotropic {HTS} from
  experimental {V-I} curves,'' \emph{Advances in Cryogenic Engineering}, vol.
  48B, pp. 1118--1125, 2002.

\bibitem{McKinnon:SIAMJO98}
K.~McKinnon, ``{Convergence of the Nelder--Mead Simplex Method to a
  Nonstationary Point},'' \emph{SIAM Journal on Optimization}, vol.~9, no.~1,
  pp. 148--158, 1998.

\bibitem{Pardo:SST11}
E.~Pardo, M.~Vojen{\v c}iak, F.~G{\"o}m{\"o}ry, and J.~{\v S}ouc,
  ``{Low-magnetic-field dependence and anisotropy of the critical current
  density in coated conductors},'' \emph{Superconductor Science and
  Technology}, vol.~24, p. 065007, 2011.

\bibitem{Zhang:JAP12}
M.~Zhang, J.-H. Kim, S.~Pamidi, M.~Chudy, W.~Yuan, and T.~A. Coombs, ``{Study
  of second generation, high-temperature superconducting coils: Determination
  of critical current},'' \emph{Journal of Applied Physics}, vol. 111, no.~8,
  p. 083902, 2012.

\bibitem{Rostila:PhysC07}
L.~Rostila, J.~Lehtonen, and R.~Mikkonen, ``Self-field reduces critical current
  density in thick ybco layers,'' \emph{Physica C}, vol. 451, pp. 66--70, 2007.

\bibitem{Gomory:SST06b}
F.~G{\"o}m{\"o}ry and B.~Klin{\v c}ok, ``{Self-field critical current of a
  conductor with an elliptical cross-section},'' \emph{Superconductor Science
  and Technology}, vol.~19, pp. 732--737, 2006.

\bibitem{Vojenciak:SST11}
M.~Vojen{\v c}iak, F.~Grilli, S.~Terzieva, W.~Goldacker, M.~Kov{\'a}c{\v o}va,
  and A.~Kling, ``Effect of self-field on the current distribution in
  {Roebel}-assembled coated conductor cables,'' \emph{Superconductor Science
  and Technology}, vol.~24, p. 095002, 2011.

\bibitem{Goldacker:SST09}
W.~Goldacker, A.~Frank, A.~Kudymow, R.~Heller, A.~Kling, S.~Terzieva, and
  C.~Schmidt, ``{Status of high transport current ROEBEL assembled coated
  conductor cables},'' \emph{Superconductor Science and Technology}, vol.~22,
  p. 034003, 2009.

\bibitem{Kario:SST13}
A.~Kario, M.~Vojenciak, F.~Grilli, A.~Kling, B.~Ringsdorf, U.~Walschburger,
  S.~I. Schlachter, and W.~Goldacker, ``{Investigation of a Rutherford cable
  using coated conductor Roebel cables as strands},'' \emph{Superconductor
  Science and Technology}, vol.~26, no.~8, p. 085019, 2013.

\bibitem{Grilli:TAS14b}
F.~Grilli, V.~M.~R. Zermeno, E.~Pardo, M.~Vojenciak, J.~Brand, A.~Kario, and
  W.~Goldacker, ``{Self-Field Effects and AC Losses in Pancake Coils Assembled
  From Coated Conductor Roebel Cables},'' \emph{IEEE Transactions on Applied
  Superconductivity}, vol.~24, no.~3, p. 4801005, 2014.

\bibitem{Ampacity_website}
\BIBentryALTinterwordspacing
Ampacity cable project. [Online]. Available: \url{http://www.rwe-ampacity.com/}
\BIBentrySTDinterwordspacing

\bibitem{Elschner:EUCAS13}
S.~Elschner, E.~Demencik, A.~Kudymow, W.~Goldacker, F.~Grilli, M.~Noe,
  S.~Strau{\ss}, M.~Vojenciak, V.~Zermeno, B.~Douine, and M.~Stemmle,
  ``{Experimental setup of a superconducting power transmission cable for
  AC-loss investigations under controlled current distribution},'' in
  \emph{$\rm 11^{th}$ European Conference on Applied Superconductivity, Genoa,
  Italy, 15-19 September 2013. Talk 1A-LS-O6}, 2013.

\bibitem{Siahrang:TAS10}
M.~Siahrang, F.~Sirois, D.~N. Nguyen, S.~Babic, and S.~P. Ashworth, ``{Fast
  Numerical Computation of Current Distribution and AC Losses in Helically
  Wound Thin Tape Conductors: Single-Layer Coaxial Arrangement},'' \emph{IEEE
  Transactions on Applied Superconductivity}, vol.~17, no.~6, pp. 2381--2389,
  2010.

\bibitem{Brambilla:SST07}
R.~Brambilla, F.~Grilli, and L.~Martini, ``{Development of an edge-element
  model for AC loss computation of high-temperature superconductors},''
  \emph{Superconductor Science and Technology}, vol.~20, no.~1, pp. 16--24,
  2007.

\bibitem{Thakur:SST11b}
K.~Thakur, A.~Raj, E.~Brandt, and S.~Saastry, ``Frequency-dependent critical
  current and transport ac loss of superconductor strip and roebel cable,''
  \emph{Superconductor Science and Technology}, vol.~24, p. 065024, 2011.

\bibitem{Zermeno:JAP13}
V.~M.~R. Zermeno, A.~B. Abrahamsen, N.~Mijatovic, B.~B. Jensen, and M.~P.
  Soerensen, ``{Calculation of AC losses in stacks and coils made of second
  generation high temperature superconducting tapes for large scale
  applications},'' \emph{Journal of Applied Physics}, vol. 114, p. 173901,
  2013.

\end{thebibliography}
\end{document}